\documentclass[aps,prl,twocolumn,epsfig]{revtex4}
\usepackage{graphicx}

\begin{document}
\tighten

\title{Disordering to Order: de Vries behavior from a Landau theory 
for smectics}

\author{Karl Saunders*,  Daniel Hernandez, and Staci
Pearson,}

\address{Deptartment of Physics, California Polytechnic State 
University, San Luis
Obispo, CA 93407, USA}

\author{John Toner}
\address{$^2$Department of Physics and Institute of Theoretical Science,
University of Oregon, Eugene, OR 97403, USA}

\date{\today}

\begin{abstract}

We show that Landau theory for the isotropic ($I$), nematic ($N$), smectic $A$, and smectic $C$ phases  generically, but not ubiquitously, implies ``de Vries'' behavior. I.e., a continuous $AC$ transition can occur with little layer contraction; the birefringence decreases as temperature $ T $ is lowered above this transition,  and increases again below the transition.  This de Vries behavior occurs in models with  unusually small orientational order, and is preceded by a first order $I-A$ transition. A first order $AC$  transition with elements of de Vries behavior can also occur. These results correspond well with  experimental work to date.
\end{abstract}


\maketitle

Recently,   an unusual new class of liquid crystals  known as ``de  Vries smectics" \cite{Spector, Giesselmann, Lagerwall, Selinger,  Clark, Panarina, Reihmann, Huang, Krueger, Hayashi} has drawn interest. They possess two defining features  \cite{Diele}. Firstly, there is little change with temperature $T$ of the layer spacing $d(T)$ in the C phase, in contrast to the rapid geometrical contraction $d(T)\propto \cos \theta (T)$ expected if the molecules tilt by a strongly  temperature angle $\theta(T)$. Secondly, the $A$ phase birefringence shows a nonmonatonic temperature dependence \cite{YuriPrivate}, initially increasing as $T$ is lowered, then decreasing as the $AC$ transition is approached. This is the first example known to us of {\it decreasing} order as  a lower symmetry
phase is approached. Generally, de Vries smectics exhibit the phase sequence $I-A-C$, without a
nematic phase, and an $AC$ transition that is second order, although in some it is  weakly first order  \cite{Huang}.

The generally  accepted picture of these materials is de Vries'  diffuse  cone model \cite{deVries}, which  says that   as the $AC$ transition is approached from the $A$ phase, the molecules ``pre-tilt'' , but in azimuthally random directions (hence reducing orientational order), so that there is no long range order in the tilting. Upon entering the $C$ phase the molecules azimuthally order without the significant layer contraction that occurs in conventional smectics whose molecules tilt at the $AC$ transition\cite{deVries, Leadbetter, Takanishi, Selinger, Clark, Giesselmann, Krueger}.

\begin{figure}
\includegraphics[angle=270,scale=0.5]{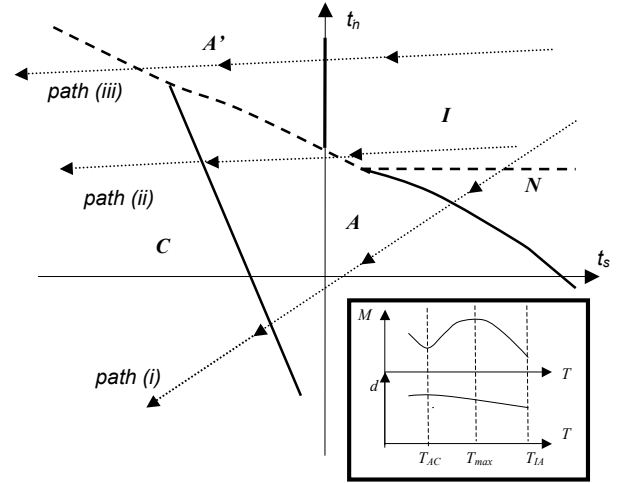}
\caption{The phase diagram in $t_s$-$t_n$ space for the  $I$, $N$, $A$, $C$ phases. First and second order phase boundaries are shown as dashed and solid lines respectively.Three decreasing  temperature paths from the $I$  to $C$ phase are shown. Path (i) corresponds to a conventional material that does not exhibit de Vries behavior.  Path (ii) corresponds to a material exhibiting de  Vries behavior and a second order $AC$ transition. Path (iii) leads to a first order $AC$ transition with elements of de Vries behavior. The inset shows the qualitative dependence on temperature of birefringence ($M$) and layer spacing ($d$).}
\label{Phase Diagram}
\end{figure}

In this Letter we show that in a complete, nonchiral Landau mean field theory for the isotropic ($I$), nematic ($N$), $A$ and $C$ phases, in which all three order parameters (orientational, layering, azimuthal tilt) and the layer spacing are coupled,  de Vries behavior occurs in a finite fraction of
parameter space (namely, for unusually weak coupling between layering and orientational order), while other regions exhibit conventional behavior. The mean field phase diagram for our model is shown in Fig. \ref{Phase Diagram}. Here $t_s$ and $t_n$ are Landau theory parameters  that control layering, and orientational order, respectively.

Note that this phase diagram   predicts  {\it two}, distinct smectic $A$  phases of identical symmetry,  denoted $A$ and $A'$, separated by a first order phase transition. While first order transitions between two smectic $A$ phases due to competition between two different layer spacings\cite{Barois} have been predicted, our result shows that even without such competition, $A-A$ transitions occur quite naturally,
and should be far more common than was previously thought.

Any experiment in which temperature is varied at fixed concentration corresponds to a particular path   through this phase diagram. As usual in Landau theory, we assume throughout this paper that $t_s$ and $t_n$ are monotonically increasing functions of temperature; hence, as temperature is lowered, one moves monotonically from upper right to lower left in Fig. \ref{Phase Diagram}. Three qualitatively distinct paths of this type are shown.  Path (i) is a typical path for  a material that does {\it not} display de Vries behavior; along it, both $t_s$ {\it and} $t_n$ both depend strongly on temperature. Paths (ii) and (iii) correspond to de Vries behavior with 2nd and 1st order $AC$ and $A'C$ transitions, respectively. Both paths have strongly varying $t_s$ and weakly varying $t_n$; i.e,  $t_n$ is virtually athermal. This would be the case if the $IN$ transition is driven by a steric mechanism for which $t_n$ depends strongly on concentration and weakly on $T$. We find that de Vries behavior occurs in this case, for sufficiently weak coupling between layering and orientational orders.

This restriction to nearly horizontal paths implies that de Vries systems should  very rarely  exhibit an $N$ phase between the $A$ and $C$ phases, since, to cross the $IN$ boundary, a nearly horizontal path in Fig. \ref{Phase Diagram} would have to be ``fine tuned'' to start very close to the boundary. The most likely paths to see de Vries behavior are those like (ii) or (iii), showing phase sequence $I-A-C$ or $I-A'-C$. This phase sequence is in good agreement with experimental work to date.

In this Letter we focus  on path (ii) and briefly  discuss path (iii) at the end. The inset in Fig. \ref{Phase
Diagram} shows our predictions for the layer spacing $d(T)$ and birefringence $M(T)$ as $T$ is varied along path (ii). The {\it increase} of the layer spacing  in the $A$ phase as the  $AC$ transition
is approached, though contrary to the de Vries picture of ``pre-tilting'' in the $A$ phase, is seen experimentally \cite{Lagerwall}.

In the $A$-phase, our Landau theory predicts:
\begin{mathletters}
\begin{eqnarray}
M_A(T) &=& M_{max} -M_2
\bigg[\frac{t_s(T)-t_s(T_{max})}{t_s(T_{max})}\bigg]^2,
\label{MA1}
\end{eqnarray}
\end{mathletters}
where $M_{max}$ , $M_2$ and $T_{max}$ are positive constants. For a finite range of  Landau   parameters, on path (ii) $T_{IA}<T_{max}<T_{AC}$, where $T_{IA}$ and $T_{AC}$ are  the $IA$ and $AC$ transition temperatures, respectively, and  $M$ has a maximum {\it within} the $A$ phase.  Hence,
$M(T)$ is nonmonatonic (the second defining feature of de Vries behavior). If $t_s$ is linear in $T$ in the
$A$-phase (as expected for  small $T_{IA} - T_{AC}$), then $M(T)$ will be {\it perfectly} parabolic in $T$.

Near  the $AC$ transition within the $C$ phase  the critical temperature dependences of $M$ and the tilt angle $\theta(T)$ predicted by our Landau theory are: $M$ linear in $T$, and $\theta \propto (T_{AC} - T)^{\frac{1}{2}}$. When fluctuation effects are included, we expect\cite{Grinstein}  $\theta \propto (T_{AC} - T)^{\beta}$, where $\beta \approx 0.35$ is the order parameter critical exponent for the 3D XY model. The layer spacing scales with temperature according to:
\begin{mathletters}
\begin{eqnarray}
d=d_0+a(T_{AC} - T)+b(T_{AC} - T)^3\;,
\label{d(T)}
\end{eqnarray}
\end{mathletters}
where $d_0$ is the value of the layer spacing {\it at} the $AC$ transition and $a$ and $b$ are constants that depend on the   Landau theory parameters . Clearly, if $a$ is sufficiently small, which we find is the case in the $C$ phase for sufficiently weak coupling between layering and orientational order, the layer spacing shows very little variation with temperature near $T_{AC}$. Significantly, we find that in cases like path (ii), the criterion for de Vries behavior of $d(T)$ differs from that for $M(T)$. Hence, we predict that some systems will exhibit de Vries behavior of the layer spacing, but {\it not} of the birefringence, and others the reverse.

While Landau  theory predicts that the $I-A'$ transition is continuous, it is known\cite {Brazovskii}  that fluctuations {\it always} drive the $I-A'$ transition first order, albeit only  {\it weakly} so  if fluctuations 
are small. Fluctuation effects will also shift the the positions of {\it all} of the transitions we've found. We expect, however, that the topology and essential geometry of the phase diagram Fig. \ref{Phase
Diagram} should occur in real systems. The only qualitative difference we expect is that the $I-A-A'$ critical end point (CEP) predicted by Landau theory will be replaced by an $I-A-A'$ {\it triple} point.
In other regions of parameter space, our model  has an $NAC$  point; we will discuss this elsewhere\cite{Saunders}.

In summary,  de Vries behavior emerges quite naturally from our Landau theory. Equally importantly, conventional behavior also genericlly occurs for different Landau parameters. Thus, the model can
accommodate all observed behaviors in {\it all} systems, and, in addition, predicts many new behaviors not yet seen experimentally, like the first order $A-A'$ and $A'-C$ transitions.

We will now briefly describe the formulation and analysis of our theory. A Landau theory for all four phases ($I$, $N$, $A$, $C$) must include order parameters for three types of order: uniaxial orientational order, layering order and tilt (azimuthal) order. All three are accounted for by the density 
$\rho=\rho_0+\delta \rho$ (with $\rho_0$ constant and $\delta  \rho$ spatially varying) and the usual
third rank tensor  orientational order parameter $\cal Q$. We take as our Landau free energy\cite{Brand} $F= F_{Q}+F_{\rho}+F_{c} $, where the orientational\cite{deGennes_book}($F_Q$), density ($F_\rho$), and coupling ($F_c$) free energies are  given by:
\begin{eqnarray}
F_{Q} &=& \int d^3 r \bigg[\frac{t_n Tr({\cal Q}^2)}{12}  - \frac{w Tr({\cal
Q}^3)}{18}  +
\frac{u_n (Tr({\cal Q}^2))^2}{144} \nonumber\\&+&K_n (\partial_i
Q_{jk})(\partial_i Q_{jk})\bigg] \;,
\label{H_Q}
\end{eqnarray}
\begin{equation}
F_{\rho} = \int d^3 r \bigg[\frac{t'_s}{4} \delta \rho^2 + 
\frac{u_s}{24} \delta
\rho^4  +\sum^\infty_{m =1} C_m  \delta \rho \nabla^{2m} \delta\rho  \bigg] \;,
\label{H_rho}
\end{equation}
\begin{eqnarray}
F_{c} &=& \int d^3 r \bigg[ -\frac{1}{4} Q_{ij} \sum^\infty_{m = 0} g_{1,m}
(\partial_i \rho)(\partial_j \nabla^{2m} \rho)
\nonumber\\
&+& \frac{g_2}{4} Q_{ik} Q_{kj}  (\partial_i \rho)(\partial_j \rho)
+\frac{g_3}{8} (Q_{ij}(\partial_i \partial_j \rho))^2
\nonumber\\
&+& \frac{h}{24} (Q_{ij}(\partial_i \rho)(\partial_j \rho))^2  \bigg] \;,
\label{H_c}
\end{eqnarray}
where $\partial_i$ is the spatial derivative in  the $i$th direction,  and the Einstein summation convention is implied. The constants $u_s$, $K_n$, $w$, $u_n$, $g_{1,m}, g_2, g_3$ and $h$ are positive.

We have for simplicity dropped some terms (e.g., $Tr({\cal Q}^2)\delta\rho^2$) that are actually lower order in the presumed small fields $\delta \rho$ and ${\cal Q}$ than the terms we have kept. We have verified that keeping such terms with small, but non-zero, coefficients has no qualitative effect on our results. In contrast, dropping any one of the terms we {\it have} kept is unphysical. For example, dropping the $h$ term destabilizes the C phase\cite{Brand}. We treat $t_n$ and $t'_s$ as the {\it only} parameters
that depend on temperature.

We seek the configuration of $\rho$ and ${\cal Q}$ that minimizes $F$. The configuration of $\rho$ that does so, is in the smectic phases, spatially modulated along an arbitrary direction which we choose to call $z$. It  can be rewritten in terms of the spatially constant complex translational (or layer) order
parameter $\Psi$, as $\delta\rho = \Psi e^{iqz} + \Psi^* e^{-iqz}$, with layer spacing $d=2 \pi /q$. With $\rho$ of this form, $\partial_i$ can effectively be replaced by $q \delta_{iz}$, which in turn allows us to define  $k(q^2) \equiv \sum^\infty_{m =1} C_m q^{2m}$ and $g_1(q) \equiv \sum^\infty_{m = 0} g_{1,m}
q^{2m}$. In order for smectic phases to occur, $k(q^2)$ must have a minimum at some value $q_0\neq0$ of $q$. In the absence of the coupling term $F_c$, and of fluctuations, $q_0$ is the layering wavevector the system would spontaneously select. Hence, if the coupling terms and fluctuations  are small, as we will assume, $q$ will be close to $q_0$, and we can expand $k(q^2)$ around $q_0^2$, thereby rewriting Eq.(\ref{H_rho}) as
\begin{mathletters}
\begin{eqnarray}
f_{\Psi} &=& \frac{1}{2} t_s |\Psi|^2 + \frac{1}{4} u_s |\Psi|^4 
+\frac{1}{2} K (q^2-q_0^2)^2|\Psi|^2  ,
  \label{f_Psi}
\end{eqnarray}
\end{mathletters}
where  we drop  oscillating terms, e.g., $\Psi^2 e^{2iqz}$,  whose $\int d^3 r$ vanish, and define $f_\Psi=F_{\rho}/V$, where $V$ is the volume, and $t_s\equiv t'_s +k(q_0^2)$ and $K\equiv\frac{\partial^2k(q^2)}{\partial (q^2)^2} |_{q=q_0}$.

The form of ${\cal Q}$  that that minimizes $F$ \cite{Saunders} is  spatially uniform, and  given by:
\begin{equation}
Q_{ij} = (-S+\sqrt{3}\eta)e_{1i} e_{1j}  +(-S-\sqrt{3}\eta)e_{2i} 
e_{2j} +(2S)e_{3i} e_{3j} \;,
\label{Q}
\end{equation}
where ${\bf \hat e_3} = {\bf c} + \sqrt{1-c^2}{\bf \hat z}$ is the average direction of the molecules' long axes, (i.e., the director). Here, in either smectic phase,  we chose ${\bf \hat z}$  normal to the  layers; in the $N$ and $I$ phases the direction of ${\bf \hat z}$ is arbitrary. The projection ${\bf c}$ of the director onto  the layers is the order parameter for the $C$ phase. The other two principal axes of ${\cal Q}$ are given by ${\bf \hat e_1} = {\bf \hat z} \times {\bf \hat c}$ and ${\bf \hat e_2} = \sqrt{1-c^2}{\bf \hat c} - c {\bf \hat z}$. $S$ and $\eta$ are proportional to the birefringence and biaxiality of the system, respectively.  The $A$ phase is untilted (${\bf c} = {\bf 0}$) and uniaxial ($\eta=0$), while  the $C$ phase is tilted (${\bf c} \neq {\bf 0}$) and biaxial ($\eta\neq 0$). It is  convenient to make the change of variables $S=M \cos(\alpha)$ and $\eta=M \sin(\alpha)$. In the $A$ phase,  $M$ is proportional to the birefringence.

We next minimize the free energy $F$ over the variables $M, \alpha, c, |\Psi|$ and $q$. Four qualitatively different types of minima are possible, corresponding to the four different symmetry phases ($I, N, A, C$). Specifically, the $I$ phase has $M=0; \Psi = 0$; the $N$ phase has $\Psi = 0$, ${\bf c}={\bf 0}$, and $\alpha = 0$, but $M\neq 0$; the $A$ phase has $\Psi \neq 0$ and $M\neq 0$, but ${\bf c}={\bf 0}$, and $\alpha = 0$; and the $C$ phase has all of the variables $M, \alpha, {\bf c},$ and $\Psi$   $\neq 0$. We
render minimization analytically tractable by assuming that the coupling term Eq. (\ref{H_c}) is small, and treating it perturbatively. This procedure leads to the phase diagram shown in Fig. \ref{Phase Diagram}.

Equations for the locii of the phase boundaries are given in \cite{online ref}. The minimization of our Landau free energy also leads to predictions for the temperature dependences of $M$ and $q$. We
find, in the $A$-phase,
\begin{mathletters}
\begin{eqnarray}
M_A &=& M_0(t_n) + \frac{q_0^2 \Psi _0^2}{\gamma}\left(-3g_2M_0 + 
\delta\right)\;,
\label{MA}
\\
q_A^2 &=& q_0^2 +\frac{M_0}{K}(-g_2 M_0+g'_1(q_0)q_0^2+\delta)\;,
\label{qA}
\end{eqnarray}
\end{mathletters}
where $\gamma\equiv w M_0-2t_n>0$ and $M_0= (w+\sqrt{w^2-4u_n t_n})/2u_n$ is the ``bare'' value of $M$, i.e. its value  in the absence of coupling. Likewise, $\Psi_0=\sqrt{-t_s/u_s}$ is the bare value of $\Psi$ and $g_1'(q_0)\equiv(\frac{dg_1(q)}{dq})|_{q=q_0}\geq0$. A small value for $g_1'(q_0)$ corresponds to a weak dependence on layer spacing of the coupling between layering and orientational order. For strongly $T$ dependent $t_s$ and athermal $t_n$ the quantity $\delta$ is most usefully expressed as
\begin{eqnarray}
\delta (t_s, t_n) &\equiv& \alpha(t_s-t_s^{AC})
\;,
\label{deltadef}
\end{eqnarray}
where $\alpha=(2 h q_0^2 M_0)/u_s$ and $t_s^{AC}=(u_s g_3 - \alpha h (g_1(q_0)-g_2M_0))/h$ is the value of $t_s$ where $\delta$ vanishes and the 2nd order $AC$ transition occurs. In the $A$ phase, $\delta>0$ and in the $C$ phase $\delta<0$. In the $C$ phase we find
\begin{mathletters}
\begin{eqnarray}
M_C &=& M_0(t_n) + \frac{q_0^2 \Psi _0^2}{\gamma}\left(-3g_2M_0 -
\frac{g_2\delta}{2 q_0^2 (g_3+h_1 \Psi_0^2)}\right)\;,
\label{MC}
\\
q_C^2 &=& q_{A}^2(t_s^{AC}) -\frac{g'_1(q_0)}{2K(g_3 + h\Psi_0^2)}\delta\;,
\label{qC}
\end{eqnarray}
\end{mathletters}
Finally, in the $A'$-phase, we find:
\begin{mathletters}
\begin{eqnarray}
M_A' &=& \frac{q_0^2 \Psi_0^2 g_1(q_0)}{t_n}\;,
\label{MC}
\\
q_A'^2 &=& q_0^2 \left(1 +\frac{2 g_1^2(q_0) \Psi_0^2}{K t_n}\right)
\;.
\label{qA'}
\end{eqnarray}
\end{mathletters}
These results imply de Vries behavior for both  birefringence and layer spacing. For  a nearly horizontal experimental locus like path (ii) through the $A$ phase, the $T$ dependence of $M$ (and hence 
birefringence) in Eq. (\ref{MA}) comes from the linear $t_s$-dependence of each of $\Psi_0^2$ and
$\delta$ in the correction due to the coupling of layering and orientational orders. From Eq. (\ref{MA}) we see that the de Vries behavior, i.e. non-monatonicity, of $M$ is due to a competition between the layering order, $\Psi_0^2$ and the coupling $\delta$ which increase and decrease respectively as the 
$AC$ transition is approached. This happens because, as the system moves deeper into the $A$ phase, the layering order increases, thereby augmenting the weak orientational order due to the coupling between the two. However, as the $AC$ transition (where at the director tilts away from the layer
normal) is approached, this coupling necessarily decreases, and, hence, so does $M_A$. It is straightforward to show that if $g_2 < 2 h q_0^2 t_s^{AC}/3u_s$ (i.e. if $g_2$ is sufficiently small) then 
this maximum is {\it inside} the $A$ phase. Thus we expect to see de Vries behavior in systems
with weak coupling between layering and orientational order.

As the $AC$ transition is approached within the $A$ phase, $q_A$ monatonically decreases and hence $d$ monatonically increases. This is typical of both conventional {\it and} de Vries smectics, although as discussed above, it is somewhat contrary to the diffuse cone picture. The $T$ dependence of the
layer spacing at the transition depends crucially on the size of the parameter $g_1'(q_0)$. In systems where the coupling of the layering and orientational order depends weakly on layer spacing and $g_1'$ is unusually small, the $T$-dependence of $q$ is almost flat. We have shown that if $g_1'(q_0)=0$ then the change in layer spacing scales like $(T_{AC} - T)^3$, and hence varies very weakly in the $C$ phase near the $AC$ transition. Systems with larger values of $g_1'(q_0)$ will have conventional behavior of the layer spacing. Since this de Vries behavior of $q$ depends on the smallness of a {\it different} parameter ($g'_1$) than did de Vries behavior of $M$ (which depended on $g_2$ being small), it should be possible to find systems which exhibit de Vries behavior of the layer spacing , but {\it not} de Vries behavior of the birefringence, or visa-versa.

For systems that approach the $C$ phase from the $A'$ phase, along path (iii), the birefringence will increase monatonically, and will not exhibit de Vries behavior. The birefringence jumps substantially at the transition (on the order of $M_0$). From Eq. (\ref{qA'}), we see that the $T$ dependent piece of $q_A'$ is second order in the coupling $g_1(q_0)$ which we treat perturbatively in our analysis. Thus, this $T$ dependent piece is very small in the $A$ phase. Upon entry into the C phase the $T$ dependence of the layering spacing will be weak if $g_1'(q_0)$ is small. At the $A'C$ transition there will be a jump in $q$. Eqs. (\ref{qA}), (\ref{qC}) and (\ref{qA'}) can be used to show that this jump will be small when $g_1'(q_0)$ and $g_2$ are small. For such a system a transition just above the CEP will exhibit a continuous change in tilt  angle, a weakly varying layer spacing, a substantial jump in birefringence, and a latent heat.  Elements of such an unusual transition have been experimentally observed \cite{Huang}.

The requirement of near $T$-independence of $t_n$ for de Vries behavior severely restricts the possible experimental locii in Fig. \ref{Phase Diagram} that can display such behavior: namely, nearly horizontal ones. A significantly sloped path like (i) will {\it not} exhibit de Vries behavior. In this case the growth of the ``bare'' (i.e., coupling-free) birefringence $M_0(t_n)$ as $T$ is lowered swamps the effects due to the coupling terms, and makes the behavior of both the birefringence and the layer spacing conventional. Thus, our model can accomodate {\it either} conventional behavior or de Vries behavior,
simply by changing parameters.

We acknowledge very helpful discussions with Yuri Panarin. K.S., D.H, S.P. were
sponsored by the Department of the Navy, Office of Naval Research.

* Corresponding author: ksaunder@calpoly.edu

\end{document}